\documentclass[12pt]{article}

\begin{document}

\hoffset=-0.75in
\voffset=-1.00in

\baselineskip=0.685cm
 
\title{Intrinsically Quantum-Mechanical Gravity and the Cosmological Constant Problem}

\author{Philip D. Mannheim \\Department of Physics,
University of Connecticut, Storrs, CT 06269, USA \\
e-mail: philip.mannheim@uconn.edu\\ }

\date{September 14, 2011}

\maketitle

We propose that gravity be intrinsically quantum-mechanical, so that in the absence of quantum mechanics the geometry of the universe would be Minkowski. We show that in such a situation gravity does not require any independent quantization of its own, with it being quantized simply by virtue of its being coupled to the quantized matter fields that serve as its source. We show that when the gravitational and matter fields possess an underlying conformal symmetry, the gravitational field and fermionic matter-field zero-point fluctuations cancel each other identically. Then, when the fermions acquire mass by a dynamical symmetry breaking procedure that induces a cosmological constant in such conformal theories, the zero-point fluctuations readjust so as to cancel the induced cosmological constant identically. The zero-point vacuum problem and the cosmological constant vacuum problems thus mutually solve each other. We illustrate our ideas in a completely solvable conformal-invariant model, namely two-dimensional quantum Einstein gravity coupled to a Nambu-Jona-Lasinio self-consistent fermion.

\section{Statement of the Problem}
\label{S1}

Included in the gravitational sources that are commonly used in astrophysics and cosmology are some intrinsically quantum-mechanical ones such as the electron Pauli degeneracy pressure that  stabilizes white dwarf stars and the black-body radiation energy density and pressure that contribute to cosmic expansion. As such, these sources contribute to the matter energy-momentum tensor $T_{\rm M}^{\mu\nu}$, and thus if the Einstein equations $(1/\kappa_4^2)G^{\mu\nu}+T_{\rm M}^{\mu\nu}=0$ are to be  treated as operator identities, they would require $G^{\mu\nu}$ to be quantum-mechanical too. But since radiative corrections to quantum Einstein gravity are not renormalizable, by hand one instead posits that the Einstein equations are to be understood as being of the semi-classical form $(1/\kappa_4^2)G_{\rm CL}^{\mu\nu}+\langle Q|T_{\rm M}^{\mu\nu}|Q\rangle=0$, with a classical $G_{\rm CL}^{\mu\nu}$ coupling to a c-number matrix element of $T_{\rm M}^{\mu\nu}$ in the quantum state $|Q\rangle$ of interest. However, since the quantum-mechanical $T_{\rm M}^{\mu\nu}$ involves products of fields at the same point, its matrix elements are not finite, and thus even though gravity couples to energy and not energy difference, in addition one equally by hand subtracts off the divergent zero-point vacuum part, to yield
\begin{equation}
\frac{1}{\kappa_4^2}G_{\rm CL}^{\mu\nu}+\langle Q|T^{\mu\nu}_{\rm M}|Q \rangle - \langle \Omega |T^{\mu\nu}_{\rm M}|\Omega \rangle_{\rm DIV}=0
\label{E1}
\end{equation}
where $|\Omega \rangle$ is the matter field vacuum. It is in the generic form (\ref{E1}) (as augmented by any classical $T^{\mu\nu}_{\rm CL}$ that might also be present) that applications of standard gravity are conventionally made.\footnote{Apart from the prescription given in (\ref{E1}), there are various other prescriptions that can be used to render quantum matter field matrix elements finite, with dimensional regularization being perhaps the most popular, especially for matter fields propagating in a curved background geometry. For the lowest order in $\hbar$ purposes of this paper, we will only need to evaluate matrix elements of $T^{\mu\nu}_{\rm M}$ in a flat background, and can thus use flat spacetime regularization procedures such as Pauli-Villars regulation. However, no matter what regularization prescription one might use, each such prescription represents a departure from the requirement that gravity couple to the energy-momentum tensor itself rather than to some modified variant thereof. In the present paper we will show that we are able to avoid the need for making any such modification by having quantum gravity do the regulation instead.} Thus, for a generic set of oscillators with Hamiltonian $H=\sum (a^{\dagger}a+1/2)\hbar \omega $, in taking one-particle matrix elements in states of the form $a^{\dagger}|\Omega\rangle$ one keeps the contribution of the $a^{\dagger}a\hbar \omega $ term and ignores the $\hbar \omega /2$ term. However, even if one does render the vacuum quantity $\langle \Omega|T^{\mu\nu}_{\rm M}|\Omega \rangle - \langle \Omega |T^{\mu\nu}_{\rm M}|\Omega \rangle_{\rm DIV}$ finite this way, the Lorentz invariance of the vacuum state still permits this quantity to be of the form  $-\Lambda g^{\mu\nu}$, with the theory thus having to possess a cosmological constant $\Lambda$ associated with the essentially uncontrollable finite part of  $\langle \Omega |T^{\mu\nu}_{\rm M}|\Omega \rangle$. Moreover, as a cooling universe goes through a cosmological phase transition, not only is an additional, potentially enormous, contribution to $\Lambda$ then induced, in addition new zero-point infinities are induced in $\langle Q|T^{\mu\nu}_{\rm M}|Q \rangle$ due to mass generation ($\omega =|\bar{k}|$ being replaced by $\omega=(k^2+m^2/\hbar^2)^{1/2}$), with a one-time subtraction term $\langle \Omega |T^{\mu\nu}_{\rm M}|\Omega \rangle_{\rm DIV}$ not being able to cancel all infinities or readily control the finite part of $\langle Q|T^{\mu\nu}_{\rm M}|Q \rangle - \langle \Omega |T^{\mu\nu}_{\rm M}|\Omega \rangle_{\rm DIV}$ at temperatures both above and below the transition temperature. The challenge to standard gravity then is to naturally recover (\ref{E1}) or some alternately regularized expression starting from a fundamental quantum gravitational theory in a way that would both clarify the nature of the subtraction procedure and naturally lead to the small value for $\Lambda$ that the theory phenomenologically requires. Since this challenge has yet to be met, in this paper we shall propose an alternate approach, one in which the difficulties associated with (\ref{E1}) are by-passed by not having an equation such as (\ref{E1}) appear at all.

To achieve this specific objective we will need to be able to construct a quantum gravitational theory that is consistent and renormalizable, so that we will then be able to use gravity itself to cancel the matter field zero-point fluctuations. Thus once one has a consistent quantum gravity theory, one then has controllable gravitational zero-point fluctuations that are available to effect the needed cancellations. Generically, if we define the action of the universe to be of the form $I_{\rm UNIV}=I_{\rm GRAV}+I_{\rm M}$, then on defining the functional variation of each one of these terms with respect to the metric to be its associated energy-momentum tensor, stationarity  with respect to the metric then yields the condition
\begin{equation}
T^{\mu\nu}_{\rm UNIV}=T^{\mu\nu}_{\rm GRAV}+T^{\mu\nu}_{\rm M}=0.
\label{E2}
\end{equation}
In theories in which the full gravitational plus matter action is renormalizable, the vanishing, and thus finiteness, of  the total $T^{\mu\nu}_{\rm UNIV}$ of the universe would survive radiative corrections and serve as an operator identity. It would thus hold in any state, and immediately lead to vacuum cancellation in the form $\langle \Omega|T^{\mu\nu}_{\rm GRAV}|\Omega \rangle+\langle \Omega|T^{\mu\nu}_{\rm M}|\Omega \rangle=0$, with the zero-point contributions of the gravitational and matter fields identically canceling each other, and with each field serving to regulate the other's divergences. Moreover, in the event of a change in vacuum to some spontaneously broken vacuum $|S\rangle$, the stationarity condition would continue to hold in the form $\langle S|T^{\mu\nu}_{\rm GRAV}|S \rangle+\langle S|T^{\mu\nu}_{\rm M}|S\rangle=0$, and thus while there would now be both mass generation and cosmological constant generation, all the various zero-point contributions would have to readjust in precisely the manner needed in order to continue to maintain the overall cancellation. The solution to the cosmological constant problem then is to treat the cosmological constant term in conjunction with the zero-point fluctuations, and in this paper we will show how this explicitly happens in a very simple solvable model. To contrast this approach with one based on (\ref{E1}), we see that in (\ref{E1}) the zero-point fluctuations are discarded before one even begins to tackle the cosmological constant problem, while in (\ref{E2}) the zero-point fluctuations play the central role. Finally, if (\ref{E2}) does hold as an operator identity, then its matrix elements in states with an indefinite number of gravitational quanta will lead to a macroscopic c-number gravitational theory that will serve as the associated classical gravity theory, in complete accord with the way one is able to transit from quantum to classical electrodynamics by taking matrix elements in states with an indefinite number of photons. 

In analyses based on (\ref{E1}), we note that already in flat spacetime the quantity $\langle \Omega|T^{\mu\nu}_{\rm M}|\Omega \rangle$ will possess zero-point contributions, while the quantity $\langle \Omega|T^{\mu\nu}_{\rm M}|\Omega \rangle - \langle \Omega |T^{\mu\nu}_{\rm M}|\Omega \rangle_{\rm DIV}$ will possess a $-\Lambda \eta^{\mu\nu}$ term where $\eta^{\mu\nu}$ is the Minkowski metric. With such vacuum terms occurring even in the absence of gravity, on expanding $G^{\mu\nu}_{\rm CL}$ as a power series in Newton's constant $\kappa_4^2$, we see that  gravity can only respond to these vacuum contributions but not control them. Indeed, it is precisely because of issues like this that the cosmological constant problem has proven to be so hard to solve, with it being very difficult for gravity to solve a problem that it is not responsible for. Moreover, if fundamental scalar Higgs fields exist, then $T^{\mu\nu}_{\rm M}$ will even contain a classical piece as well, to provide yet another term over which gravity would have no control.

In order to give gravity control of the problems that afflict it, we thus propose to put gravity on an equal footing with matter by expanding the metric not as a power series in the gravitational coupling constant but as a power series in Planck's constant instead. Additionally, we propose that in the absence of quantum mechanics there would be no curvature at all with all the mass and length scales needed to characterize spacetime curvature being intrinsically quantum-mechanical, so that in the absence of quantum mechanics the geometry would be Minkowski. Then, with curvature only occurring in the presence of $\hbar$ there can be no classical contributions to $T^{\mu\nu}_{\rm M}$, with any mass-generating symmetry breaking needing to be effected via dynamical fermion condensates rather than by fundamental Higgs fields.\footnote{We thus propose that no fundamental scalar fields exist in nature at all, with  the fundamental Lagrangian having to consist of fermions, gauge bosons and the metric tensor alone. However, just as in the Nambu-Jona-Lasinio four-Fermi model [Y.~Nambu and G.~Jona-Lasinio, Phys.~Rev.~{\bf 122}, 345 (1961)], a model that expressly possesses no fundamental scalar fields, there can of course still be dynamical fermion anti--fermion scalar bound states. In such a situation the Higgs field that is to appear in the Weinberg-Salam theory is to be understood not as a fundamental field at all, but as a c-number matrix element of a fermion bilinear composite in a degenerate vacuum state.} Thus we propose that  $T^{\mu\nu}_{\rm M}$ and $T^{\mu\nu}_{\rm GRAV}$ both be intrinsically quantum-mechanical with neither containing any intrinsic classical contributions whatsoever. As we will see, this will lead us to a natural resolution of the vacuum energy problem, and as a bonus we will find that we do not need to quantize gravity independently. Rather, once $T^{\mu\nu}_{\rm M}$ is quantized,  $T^{\mu\nu}_{\rm GRAV}$ will be quantized simply by virtue of its being coupled to $T^{\mu\nu}_{\rm M}$ in (\ref{E2}). Moreover, with there being no intrinsic classical gravity, one no longer needs to address the issue of how quantization might affect some given classical gravitational configuration.

In order to implement the above proposal we require two things,  a theory of gravity in which there are no non-trivial classical solutions and a theory of gravity in which quantum radiative corrections are under control. Remarkably, both of these requirements are met if we endow both gravity and its matter sources with an underlying local conformal invariance. For gravity this would mean conformal gravity as based on the Weyl action $I_{\rm W}=-\alpha_g\int d^4x (-g)^{1/2}C_{\lambda\mu\nu\kappa}C^{\lambda\mu\nu\kappa}$ (see e.g. \cite{R1}) or its conformal supergravity extension, while for the matter fields it would mean the standard $SU(3)\times SU(2)\times U(1)$ theory of strong, electromagnetic and weak interactions as constructed with massless fermions and gauge bosons, but without any  fundamental double-well Higgs potential with its non-conformal invariant tachyonic mass. In such a theory there are no fundamental scales at the level of the Lagrangian, with all scales being generated quantum-mechanically either via the non-vanishing of quantum-mechanical commutation relations \footnote{The generic equal time commutation relation $[\phi(\bar{x},t),\pi(\bar{x}^{\prime}, t)]=i\hbar \delta^3(\bar{x}-\bar{x}^{\prime})$ is a non-linear relation that introduces a scale $\delta^3(\bar{x}-\bar{x}^{\prime})$ everywhere on a spacelike hypersurface. Since we can set $\delta^3(\bar{x}-\bar{x}^{\prime})=(1/8\pi^3)\int d^3k \exp(i\bar{k}\cdot (\bar{x}-\bar{x}^{\prime}))$, this is equivalent to introducing a complete basis of momentum space eigenmodes. And indeed, it is this very set of modes that gives rise to the zero-point energy density and pressure of a quantized field.} or via dynamical mass generation by fermion bilinear condensates. When conformal gravity is treated classically, we see that the only allowed solutions to the gravitational equations of motion must be Minkowski, since the classical gravitational equations contain no fundamental classical length scales that could be used to parameterize any departures of the geometry from flat. At the quantum level, conformal gravity is a power-counting renormalizable since in four spacetime dimensions the gravitational coupling constant $\alpha_g$ is dimensionless, but because its equations of motion contain fourth-order derivatives it had long been thought that a quantum gravity theory based on it would not be unitary. However, on explicitly constructing the relevant Hilbert space, the theory was found \cite{R2,R3,R4,R4p} to be free of both negative norm and negative energy states, and it can thus be recognized as a viable quantum gravity theory. In \cite{R4,R4p,R5} we discuss the full conformal gravity theory itself, while in this paper we illustrate our ideas using a much simpler conformal invariant theory, namely two-dimensional Einstein gravity with action $I_{\rm GRAV}=-(1/2\kappa_2^2)\int d^2x (-g)^{1/2}R^{\alpha}_{\phantom{\alpha}\alpha}$, as two-dimensional spacetime is the spacetime in which Newton's constant is dimensionless.  In the following  it will be the very existence of  an underlying conformal symmetry that will play the central role. First, it  forbids the presence of any possible fundamental cosmological constant term at the level of the input action, thereby making conformal invariance at the level of the action an ideal starting point  to attack the cosmological constant problem. Then, with the same conformal symmetry  forcing the vanishing of the trace of the matter field energy-momentum tensor, the various vacuum contributions to it have to mutually cancel each other identically no matter how large they might individually be. In this way the cosmological constant term that is induced via dynamical mass generation is then cancelled. Conformal symmetry thus controls the cosmological constant both before and after the conformal symmetry is broken.\footnote{As discussed in more detail in \cite{R4p}, the essence of our approach here is to define the classical limit of quantum gravity as the c-number macroscopic theory obtained by taking matrix elements of the microscopic quantum field operators in states containing an indefinite number of gravitational quanta. In this way phenomena such as macroscopic gravitational waves can be generated. However, what there could not be is gravity waves that are intrinsically classical. For such an approach to succeed and emulate electrodynamics (where macroscopic classical electrodynamics emerges from microscopic quantum electrodynamics via matrix elements in states containing an indefinite number of photons), it is necessary that like quantum electrodynamics, the microscopic quantum gravity be consistent and renormalizable. Our ideas can thus work in conformal quantum gravity. However, since such a theory is conformal, it must possess no intrinsic mass scales at the level of the Lagrangian. Rather, mass scales can only be generated by a symmetry breaking procedure in which mass scales enter the theory not through the operator structure of the theory at all but through the degenerate vacuum Hilbert space states on which the operators act. Since symmetry breaking via a degenerate vacuum is intrinsically quantum-mechanical, all mass scales are intrinsically quantum-mechanical too. Thus in the absence of quantum mechanics there could be no generation of mass scales with which to characterize spacetime curvature, and the geometry would have to be Minkowski. This intricate connection between gravitational curvature and mass generation is one of the central components of the approach to the cosmological constant problem that we present here.}

\section{Zero-Point Energy Density and Zero-Point Pressure}
\label{S2}

To illustrate the nature of the vacuum issues that are involved, it is convenient to first look at the vacuum expectation value of the energy-momentum tensor $T^{\mu\nu}_{\rm M}=i\hbar \bar{\psi}\gamma^{\mu}\partial^{\nu}\psi$ of a free fermion of mass $m$ in flat, four-dimensional spacetime. With $k^{\mu}=((k^2+m^2/\hbar^2)^{1/2},\bar{k})$ it evaluates to 
\begin{equation}
\langle \Omega |T^{\mu\nu}_{\rm M}|\Omega\rangle= -\frac{2\hbar}{(2\pi)^3}\int_{-\infty}^{\infty}d^3k\frac{k^{\mu}k^{\nu}}{k^0}.
\label{E3}
\end{equation}
In (\ref{E3}) we recognize two infinite zero-point terms, one associated with $\rho_{\rm M}=\langle \Omega |T^{00}_{\rm M}|\Omega\rangle$ and the other with $p_{\rm M}=\langle \Omega |T^{11}_{\rm M}|\Omega\rangle=\langle \Omega |T^{22}_{\rm M}|\Omega\rangle=\langle \Omega |T^{33}_{\rm M}|\Omega\rangle$. However, these two terms do not have the form of a cosmological constant term since the two terms do not obey $p_{\rm M}=-\rho_{\rm M}$. And indeed, the two terms could not obey such a relationship since in the event that the fermion is massless, the associated vanishing of $k^{\mu}k_{\mu}$ would then entail that $3p_{\rm M}-\rho_{\rm M}=0$, while the trace of $-\Lambda \eta^{\mu\nu}$ is given by  the non-zero $-4\Lambda$. Given its $k^{\mu}k^{\nu}$ structure, $\langle \Omega |T^{\mu\nu}_{\rm M}|\Omega\rangle$ can be written in the form of a generic perfect fluid with a timelike fluid velocity vector $U^{\mu}=(1,0,0,0)$, viz. 
\begin{equation}
\langle \Omega |T^{\mu\nu}_{\rm M}|\Omega\rangle= (\rho_{\rm M}+p_{\rm M})U^{\mu}U^{\nu}+p\eta^{\mu\nu},~~~
\eta_{\mu\nu}\langle \Omega |T^{\mu\nu}_{\rm M}|\Omega\rangle=3p_{\rm M}-\rho_{\rm M},
\label{E4}
\end{equation}
with the fluid thus possessing both a zero-point energy density and a zero-point pressure, and with each of these quantities being divergent.\footnote{The presence of the timelike fluid 4-vector in (\ref{E4}) is due to the fact that the integration in (\ref{E3}) is over on-shell fermion modes, to thus be a three-dimensional integration and not a four-dimensional one, with the time and space components of $k^{\mu}$ being treated differently. Even though (\ref{E3}) involves terms that are infinite and thus not well-defined, we note the perfect fluid form given in (\ref{E4}) can be established by integrating over the direction of the 3-momentum vector $\bar{k}$ alone, an integration that is completely finite. The perfect fluid form for (\ref{E3}) can thus be established prior to the subsequent divergent integration over the magnitude of the momentum, with this latter integration not bringing $\langle \Omega |T^{\mu\nu}_{\rm M}|\Omega\rangle$ to the form of a cosmological constant. Even though (\ref{E3}) is not well-defined, for our purposes here the perfect fluid form given in (\ref{E4}) is a very convenient way of summarizing the infinities in (\ref{E3}) that need to be cancelled.} Since gravity couples to the full  $T^{\mu\nu}_{\rm M}$ and not just to its $(0,0)$ component, it is not sufficient to only address the vacuum energy density problem, one has to deal with the vacuum pressure as well. There are thus two vacuum problems that need to be addressed, and not just one. Moreover, since gravity couples to energy density and not to energy density difference, one is not free to remove matter field zero-point contributions by normal ordering. Rather, one needs all of the various fields in the theory to mutually cancel each others' zero-point divergences. As we will show below, conformal invariance will precisely achieve this for us.

Now as such, the existence of a vacuum perfect fluid with the above infinite energy density and pressure would appear to violate the well-known theorem that in flat spacetime the vacuum matrix element of the matter field energy-momentum tensor must take the form $\langle \Omega |T^{\mu\nu }_{\rm M}|\Omega\rangle=-\Lambda \eta^{\mu\nu}$ where $\Lambda$ is a zero or non-zero constant.\footnote{The form found in (\ref{E3}) would appear to violate a second theorem that also follows from the Lorentz invariance of a  flat spacetime vacuum, namely that the total energy of the vacuum has to be zero. However, this second theorem is incompatible with the requirement that the energy density be given by $\langle \Omega |T^{00 }_{\rm M}|\Omega\rangle=-\Lambda \eta^{00}$, since once the energy density is given by a non-zero constant $\Lambda$,  the total energy (its flat space 3-volume integral) would be infinite. Infinite total energy thus need not violate Lorentz invariance.} However, as we expressly show below, such a theorem only holds after $\langle \Omega |T^{\mu\nu }_{\rm M}|\Omega\rangle$ has been made finite. In order to make $\langle \Omega |T^{\mu\nu }_{\rm M}|\Omega\rangle$ finite, one needs to first identify its infinities, and they are summarized by the form given in (\ref{E4}). A viable theory of gravity thus has to deal with the divergences that are present in $\langle \Omega |T^{\mu\nu }_{\rm M}|\Omega\rangle$. Now while infinities in $\langle \Omega |T^{\mu\nu }_{\rm M}|\Omega\rangle$ arise because $T^{\mu\nu }_{\rm M}$ involves products of matter fields at the same point, we note that this equally true of $T^{\mu\nu }_{\rm GRAV}$ as introduced in (\ref{E2}). However from the fact that   $T^{\mu\nu }_{\rm GRAV}+T^{\mu\nu }_{\rm M}$ is zero, all the vacuum infinities in $T^{\mu\nu }_{\rm GRAV}$ and $T^{\mu\nu }_{\rm M}$ must mutually cancel, leaving $T^{\mu\nu }_{\rm UNIV}$ not only finite, but even zero.  While neither $T^{\mu\nu }_{\rm GRAV}$ or $T^{\mu\nu }_{\rm M}$ is separately well-defined each one can be written as $T^{\mu\nu }_{\rm GRAV}=(T^{\mu\nu }_{\rm GRAV})_{\rm FINITE}+(T^{\mu\nu }_{\rm GRAV})_{\rm DIV}$, $T^{\mu\nu }_{\rm M}=(T^{\mu\nu }_{\rm M})_{\rm FINITE}+(T^{\mu\nu }_{\rm M})_{\rm DIV}$. Then with the explicit cancellation $(T^{\mu\nu }_{\rm GRAV})_{\rm DIV}+(T^{\mu\nu }_{\rm M})_{\rm DIV}=0$ that we obtain, we can set $T^{\mu\nu }_{\rm UNIV}=(T^{\mu\nu }_{\rm GRAV})_{\rm FINITE}+(T^{\mu\nu }_{\rm M})_{\rm FINITE}$, with everything now being well-defined. Then with $T^{\mu\nu }_{\rm UNIV}$ being both well-defined and zero, we see that the total energy of the universe is zero.

For the purposes of actually parameterizing the divergences in (\ref{E3}) it is convenient to introduce a non-covariant momentum cut-off, with $\rho_{\rm M}$ and $p_{\rm M}$ then being given by 
\begin{eqnarray}
\rho_{\rm M}&=&-\frac{\hbar}{4\pi^2}\left(K^4 +\frac{m^2K^2}{\hbar^2} -{m^4 \over 4\hbar^4}{\rm ln}\left({4\hbar^2K^2 \over m^2}\right)+{m^4\over 8\hbar^4}\right),
\nonumber \\
p _{\rm M}&=&- \frac{\hbar}{12\pi^2}\left(K^4 -\frac{m^2K^2}{\hbar^2} +{3m^4 \over 4\hbar^4}{\rm ln}\left({4\hbar^2K^2 \over m^2}\right)-{7m^4\over 8\hbar^4}\right).
\label{E5}
\end{eqnarray}
We thus encounter a mass-independent quartic divergence and mass-dependent quadratic and logarithmic divergences. Hence, mass generation will not merely change the vacuum energy density and pressure, it will change them by infinite amounts, an effect that we will take care of below by having the mass generation be associated with an induced cosmological constant that will equally be divergent. Thus the cancellation $(T^{\mu\nu }_{\rm GRAV})_{\rm DIV}+(T^{\mu\nu }_{\rm M})_{\rm DIV}=0$ will hold in both the normal vacuum and spontaneously broken vacuum Hilbert spaces, and as we show, that will resolve the cosmological constant problem.\footnote{While the change in the vacuum energy is finite in the fundamental Higgs field case where one shifts from the local maximum to the minimum in the Higgs potential, in the dynamical fermion condensate case that occurs in the Nambu-Jona-Lasinio model the shift is infinite. In the dynamical case then, the resolution of the cosmological constant problem is tied in with the mechanism needed to make the energy-momentum tensor finite.}

If one wishes to define the integral in (\ref{E3}) via the use of a set of covariant Pauli-Villars regulator masses $M_i$ with each such regulator contributing an analog of (\ref{E3}) as multiplied by some overall factor $\eta_i$ (due to Hilbert space metric signature and/or the fermionic or bosonic nature of the regulator), the choice $1+\sum\eta_i=0$, $m^2+\sum \eta_iM_i^2=0$,   $m^4+\sum \eta_iM_i^4=0$ will not only then lead to finite regulated $\rho_{\rm REG}$ and $p_{\rm REG}$, it will give them the values
\begin{equation}
\rho_{\rm REG}=-p_{\rm REG}=- \frac{\hbar}{16\pi^2}\left(m^4{\rm ln}m^2+\sum \eta_i M_i^4{\rm ln}M_i^2\right).
\label{E6}
\end{equation}
The regulation procedure will thus make $\rho_{\rm REG}+p_{\rm REG}$ be equal to zero, just as required of a cosmological constant term, with a regulated $\langle \Omega |T^{\mu\nu }_{\rm M}|\Omega\rangle$ then behaving as $-\rho_{\rm REG}\eta^{\mu\nu}$. Thus we see that it is only the finite part of $\langle \Omega |T^{\mu\nu }_{\rm M}|\Omega\rangle$ that will behave like a cosmological constant term (though it would be a huge one if (\ref{E6}) is any indicator), while its infinite part could have a more complicated structure even if the vacuum is Lorentz invariant. Thus, as noted above, in general we have to deal not just with a vacuum energy density problem, but with a vacuum pressure problem as well. Moreover, while different choices of energy-momentum tensor can be made that all lead to the same total energy (such as the canonical one or the Belinfante one for instance), these various choices lead to differing expressions for the pressure. Thus to correctly define the pressure,  one must define the energy-momentum tensor as the functional variation with respect to the metric of a general-coordinate invariant action, just as in (\ref{E2}). 

As we see from the structure of (\ref{E5}), in a massless theory there will only be a quartically divergent term. We can cancel this term by having some additional quartic divergence with the opposite sign. Moreover, this is precisely how supersymmetry does the cancellation because the vacuum energy densities of  fermionic and bosonic fields have opposite signs. However, for this cancellation mechanism to be maintained following mass generation, the fermions and their bosonic superpartners would need to acquire the same masses (which experimentally we know not to be the case), since there would otherwise be uncanceled quadratic and logarithmic divergences. And even if these uncanceled terms are to be associated with the cut-off scale of some low energy effective theory, they would still make a huge contribution in (\ref{E1}). Thus using either supersymmetry or regulators as in (\ref{E6}), one can anticipate an eventual huge effective cosmological constant, and it has yet to be shown that this is not in fact the case in theories based on (\ref{E1}).

One remaining option is to cancel the quartic divergences of the massless fermionic theory by conformal symmetry instead, and here the needed bosonic contribution would come from the gravitational sector as per (\ref{E2}) rather than (\ref{E1}). Moreover, as we show below, this particular cancellation mechanism will not be destroyed by mass generation. However, if we try to regulate the fermion vacuum energy as in (\ref{E6}), while such regulators would not violate Lorentz invariance, their masses would violate conformal invariance and lead to conformal anomalies. To avoid any such anomalies we must not introduce any such regulation. Rather, we must have gravity itself do the cancellation just as in (\ref{E2}). Thus we need to cancel $\langle \Omega |T^{\mu\nu }_{\rm M}|\Omega\rangle_{\rm DIV}$ mode by mode so as to eliminate the need to do any integration over modes. As we will see, this is precisely what quantizing gravity will guarantee us, with (\ref{E2}) fixing the normalization of the gravity sector commutators so as to ensure that the needed cancellations provided by $\langle \Omega |T^{\mu\nu }_{\rm GRAV}|\Omega\rangle_{\rm DIV}$ explicitly occur mode by mode. Since we thus can ignore conformal anomalies, in the following we will be able to take advantage of the tracelessness of both the matter and the gravitational energy-momentum tensors that is required by their underlying conformal structure.\footnote{ With the vanishing of $T^{\mu\nu}_{\rm UNIV}$ being due to stationarity with respect to the metric, such stationarity equally guarantees the  vanishing of the trace $g_{\mu\nu}T^{\mu\nu}_{\rm UNIV}$ without any need to impose conformal invariance. Thus even though the vanishing of the individual gravity sector and matter sector traces $g_{\mu\nu}T^{\mu\nu}_{\rm GRAV}$ and $g_{\mu\nu}T^{\mu\nu}_{\rm M}$ do require conformal invariance, and even though conformal symmetry Ward identities might be violated by renormalization anomalies, the vanishing of $g_{\mu\nu}T^{\mu\nu}_{\rm UNIV}$ cannot be affected by the lack of scale invariance of regulator masses. Any anomalies in $g_{\mu\nu}T^{\mu\nu}_{\rm GRAV}+g_{\mu\nu}T^{\mu\nu}_{\rm M}$ must thus all mutually cancel each other identically, and one can thus proceed as though both of the $g_{\mu\nu}T^{\mu\nu}_{\rm GRAV}$ and $g_{\mu\nu}T^{\mu\nu}_{\rm M}$ traces are anomaly free.}

\section{Two-Dimensional Quantum Einstein Gravity}
\label{S3}

While we will study two-dimensional Einstein gravity since it very  straightforwardly embodies all of the general ideas we present here and in the four-dimensional conformal gravity study given in  \cite{R4,R4p,R5}, we should note that two-dimensional Einstein gravity is not the most general two-dimensional gravity theory that one could consider. Nor is it even the most general conformal invariant one. The Polyakov \cite{P1} bosonic string action $I_{\rm P}=-(1/4\pi \alpha^{\prime})\int d\tau d\sigma (-\gamma)^{1/2}\gamma^{ab}\eta^{\mu\nu}\partial_aX_{\mu}\partial_b X_{\nu}$ with $\gamma={\rm det}\gamma_{ab}$ is conformal invariant on the two-dimensional $\tau,\sigma$ world sheet because the $D$-dimensional spacetime coordinates $X_{\mu}$ and metric $\eta^{\mu\nu}$ are taken to have conformal weight zero. However, since the coordinates  $X_{\mu}$ carry the dimension of length, the Regge slope  $\alpha^{\prime}$ has to have dimension length squared. Thus, Einstein gravity is the most general conformal invariant theory that one can write down in two dimensions that is free of intrinsic length scales.\footnote{While beyond the scope of the present paper, if one wishes to include Grassmann spinors, then as well as the standard supersymmetric fermionic string extension of the bosonic string given in  A.~M.~Polyakov, Phys.~Lett.~B {\bf 103}, 211 (1981), one could also consider a pure Grassmann variant of the bosonic string based on the action $I_{\rm GS}=\alpha_{\rm GS}\int d\tau d\sigma (-\gamma)^{1/2}\gamma^{ab}\partial_a\bar{\theta}\partial_b \theta$ where $\theta$ is a Grassmann spinor (GS) coordinate. If these spinors are taken to have zero conformal weight, then just like the bosonic string action, the $I_{\rm GS}$ action would also be locally conformal invariant on the world sheet. In the event that the spinors have no spacetime dimension, the coefficient $\alpha_{\rm GS}$ would be dimensionless.} In this regard it is thus the two-dimensional analog of four-dimensional conformal gravity.

Now while the  Einstein-Hilbert action $I_{\rm GRAV}=-(1/2\kappa_2^2)\int d^2x (-g)^{1/2}R^{\alpha}_{\phantom{\alpha}\alpha}$ is locally conformal invariant  in two dimensions (Newton's constant $\kappa_2^2$ being dimensionless in two dimensions), this action has the property that as a classical action, it is a total divergence in any gravitational path (the two-dimensional Gauss-Bonnet theorem). Consequently the classical Einstein tensor will vanish identically for any choice of classical metric $g^{\mu\nu}(x)$ whatsoever. Moreover, with quantum correlators being given via Feynman path integration of the classical action over classical gravitational metrics, path integration is trivial, and quantum-mechanically there is no gravitational scattering. Thus the classical theory does not exist, and quantum radiative corrections do not exist either. Hence, if one is interested in constructing a two-dimensional quantum gravity theory in which there is to be gravitational scattering, one must change the dynamics, with the Polyakov bosonic string theory then precisely serving this purpose.

Despite the fact that there is no quantum gravitational scattering in two-dimensional Einstein gravity, that does not mean that the theory is completely empty. Specifically, once one requires field theory commutators to be non-vanishing, zero-point fluctuations will then occur.  However, these fluctuations are not contained within the path integral quantization procedure. Rather, the path integral only generates correlators that are built out of normal-ordered products of fields. And indeed, the path integral must only contain the finite normal-ordered piece of the vacuum energy, as it otherwise would not exist. Since we will show below that there are non-trivial zero-point fluctuations in two-dimensional quantum Einstein gravity, the very absence of quantum scattering in the theory enables us to isolate and focus on the zero-point fluctuation issue, with two-dimensional Einstein gravity not just being appropriate for our purposes here, it might possibly be unique in this regard.\footnote{In this regard gravity differs from other field theories. In all other field theories one cannot measure energy but only energy difference, so that for them one is free to normal order away the zero-point vacuum energy density. And with path integration generating correlators of products of fields that are normal-ordered, in such field theories the zero-point vacuum energy density has no physical relevance. However, the hallmark of gravity is that gravity couple to energy rather than to energy difference, and it thus does couple to the zero-point energy density. This coupling is not contained within path integral quantization, and can exist and be non-trivial even if path integration leads to a trivial S-matrix. In this paper we calculate the zero-point energy density in a theory (two-dimensional quantum Einstein gravity) in which path integration is trivial, and explicitly show that the zero-point energy density is non-zero. It is through recognizing the special status of zero-point fluctuations in the gravitational case that we are able to develop the approach to the cosmological constant problem we present here.}

In analog to (\ref{E5}), for a free massless fermion in a two-dimensional flat spacetime, a canonical quantization of the form $\{\psi_{\alpha}(x,t),\psi^{\dagger}_{\beta}(x^{\prime},t)\}=\delta(x-x^{\prime})\delta_{\alpha,\beta}$ leads to a perfect fluid form for $\langle \Omega |T^{\mu\nu}_{\rm M}|\Omega\rangle$ with
\begin{equation}
\rho_{\rm M}=p _{\rm M}=- \frac{\hbar K^2}{2\pi},
\label{E7}
\end{equation}
and a two-dimensional trace $p_{\rm M}-\rho_{\rm M}$ that vanishes. The task for two-dimensional quantum Einstein gravity is thus to cancel this quadratic divergence. To see how this can be achieved, we note that in generic relations such as $A\partial_{\mu}B+B\partial_{\mu}A =\partial_{\mu}(AB)+[B,\partial_{\mu}A]$, because of ordering, a function that would be a total divergence classically need not be one quantum-mechanically. Hence in the presence of quantum ordering two-dimensional quantum Einstein gravity need no longer be trivial. To take ordering into account, we need to specify a choice of ordering, and in order to enforce symmetry of the Ricci tensor, we once and for all define geometric tensors according to the ordering sequencing $R_{\lambda\mu\nu\kappa}=(1/2)[\partial_{\kappa}\partial_{\mu}g_{\lambda\nu}  -\partial_{\kappa}\partial_{\lambda} g_{\mu\nu}  
-\partial_{\nu}\partial_{\mu}g_{\lambda\kappa} 
+\partial_{\nu}\partial_{\lambda}g_{\mu\kappa})
+g_{\eta\sigma}(\Gamma^{\eta}_{\nu\lambda}\Gamma^{\sigma}_{\mu\kappa}
-\Gamma^{\eta}_{\kappa\lambda}\Gamma^{\sigma}_{\mu\nu}]$, 
 $\Gamma^{\alpha}_{\mu\kappa}=(1/2)g^{\alpha\beta}(\partial_{\mu}g_{\beta\kappa}+\partial_{\kappa}g_{\beta\mu}-\partial_{\beta}g_{\mu\kappa})$, $R_{\mu\kappa}=(1/2)[g^{\nu\lambda}R_{\lambda\mu\nu\kappa}+g^{\nu\lambda}R_{\lambda\kappa\nu\mu}]$, $R^{\alpha}_{\phantom{\alpha}\alpha}=g^{\mu\kappa}R_{\mu\kappa}$, and $G_{\mu\kappa}=R_{\mu\kappa} -(1/2)g_{\mu\kappa}R^{\alpha}_{\phantom{\alpha}\alpha}$. 
 
If we perturb to second order around flat spacetime according to $g_{\mu\nu}=\eta_{\mu\nu}+h_{\mu\nu}$, $g^{\mu\nu}=\eta^{\mu\nu}-h^{\mu\nu}+h^{\mu}_{\phantom{\mu}\sigma}h^{\sigma \nu}$, then we find that the first order $G_{\mu\nu}(1)$ vanishes identically (as it of course must since there is no ordering issue in first order and $G_{\mu\nu}$ already vanishes classically). However, for our choice of ordering, in second order we obtain 
\begin{eqnarray}
G_{00}(2)&=&\frac{1}{4}[\partial_0h_{00},\partial_1h_{01}]+\frac{1}{4}[\partial_1h_{11},\partial_0h_{01}] +\frac{1}{8}[\partial_0h_{11},\partial_0h_{00}] +\frac{1}{8}[\partial_1h_{00},\partial_1h_{11}]=G_{11}(2),
\nonumber\\
G_{01}(2)&=&\frac{1}{8}[\partial_0h_{00},\partial_1h_{00}] +\frac{1}{8}[\partial_1h_{11},\partial_0h_{11}]+\frac{1}{4}[\partial_1h_{00},\partial_1h_{01}]+\frac{1}{4}[\partial_0h_{11},\partial_0h_{01}],
\label{E8}
\end{eqnarray}
with the two-dimensional $G^{\mu\nu}(2)$ automatically being traceless, just as it should be in a conformal theory. As we thus see, the quantum $G_{\mu\nu}(2)$ is given by a set of commutator terms, terms that would vanish classically but not quantum-mechanically, to thus make the quantum theory non-trivial despite the triviality of the classical theory. Given (\ref{E8}) we can evaluate the various components of the covariant derivative of $G^{\mu\nu}(2)$, to obtain
\begin{eqnarray}
\partial_{\mu} G^{\mu}_{\phantom{\mu} 0}(2)&=&\frac{1}{4}[\nabla^2h_{00},\partial_1h_{01}]-\frac{1}{4}[\nabla^2h_{01},\partial_1h_{11}]
-\frac{1}{4}[\partial_0\partial_1h_{01},\partial_0h]-\frac{1}{8}[\partial_0\partial_1(h_{00}+h_{11}),\partial_1h]
\nonumber\\
&&+\frac{1}{4}[\partial_1^2h_{01},\partial_1h]
+\frac{1}{8}[\nabla^2h,\partial_0h_{00}]
+\frac{1}{8}[\partial_1^2h_{11},\partial_0h]
+\frac{1}{8}[\partial_0^2h_{00},\partial_0h],
\nonumber\\
\partial_{\mu} G^{\mu}_{\phantom{\mu} 1}(2)&=&\frac{1}{4}[\nabla^2h_{11},\partial_0h_{01}]-\frac{1}{4}[\nabla^2h_{01},\partial_0h_{00}]
-\frac{1}{4}[\partial_0\partial_1h_{01},\partial_1h]-\frac{1}{8}[\partial_0\partial_1(h_{00}+h_{11}),\partial_0h]
\nonumber\\
&&+\frac{1}{4}[\partial_0^2h_{01},\partial_0h]
-\frac{1}{8}[\nabla^2h,\partial_1h_{11}]
+\frac{1}{8}[\partial_0^2h_{00},\partial_1h]
+\frac{1}{8}[\partial_1^2h_{11},\partial_1h],
\label{E9}
\end{eqnarray}
where $\nabla^2=-\partial_0^2+\partial_1^2$, $h=\eta^{\mu\nu}h_{\mu\nu}=-h_{00}+h_{11}$. Since $G_{\mu\nu}(1)$ vanishes trivially, there is no first-order equation of motion that would force $\partial_{\mu} G^{\mu\nu}(2)$ to vanish identically. Consequently, the Bianchi identity of classical gravity does not automatically have to hold quantum-mechanically. Nonetheless, because of the matter field wave equation, the matter field energy-momentum tensor is covariantly conserved. Thus from the quantum-mechanical field equation 
\begin{equation}
\frac{1}{\kappa_2^2}G^{\mu\nu}+T^{\mu\nu}_{\rm M}=0,
\label{E10}
\end{equation}
it follows that $\partial_{\mu} G^{\mu\nu}(2)$ does vanish after all. As we thus see, unlike the standard classical situation, in the quantum theory  $G^{\mu\nu}(2)$ is only covariantly conserved on the stationary path and not on the arbitrary one.

Since $T^{\mu\nu}_{\rm M}$ is of order $\hbar$ in (\ref{E7}), to satisfy (\ref{E10}) to this order we must take $h_{\mu\nu}$ to be of order $h^{1/2}$, with (\ref{E10}) then fixing the gravitational commutators in $G^{\mu\nu}(2)$ to be of order $\hbar$. It is thus the quantization of the matter field that forces the quantization of the gravitational field with $G^{\mu\nu}$ not being able to vanish once $T_{\rm M}^{\mu\nu}$ does not. Through order $\hbar$ we can take $T^{\mu\nu}_{\rm M}$  to have the value it would have in flat space, with curvature corrections to it only appearing in higher order in $\hbar$. However, the quantum $G^{\mu\nu}$ is non-trivial already in order $\hbar$. Finally, with the vanishing of  $\partial_{\mu} G^{\mu\nu}(2)$ as now enforced by the vanishing of  $\partial_{\mu} T_{\rm M}^{\mu\nu}(2)$, from (\ref{E10}) we see that the components of $h_{\mu\nu}$ can be taken to obey
\begin{equation}
\nabla^2 h_{00}=0,\qquad \nabla^2 h_{01}=0,\qquad \nabla^2 h_{11}=0,\qquad h=-h_{00}+h_{11}=0,
\label{E11}
\end{equation}
to thus obey a massless wave equation after all, with the trace of $h_{\mu\nu}$ vanishing just as is to be expected in a conformal theory.

Given (\ref{E11}), we can expand the quantum fields in a complete basis of plane waves with $k^{\mu}=(\omega_k,k)$ where $\omega_k=|k|$. We thus introduce creation and annihilation operators and set 
\begin{eqnarray}
h_{00}(x,t)&=&\kappa_2\hbar^{1/2}\int \frac{dk}{(2\pi)^{1/2}(2\omega_k)^{1/2}}\left[A(k)e^{i(kx-\omega_kt)}+C(k)e^{-i(kx-\omega_kt)}\right]=h_{11}(x,t),
\nonumber\\
h_{01}(x,t)&=&\kappa_2\hbar^{1/2}\int \frac{dk}{(2\pi)^{1/2}(2\omega_k)^{1/2}}\left[B(k)e^{i(kx-\omega_kt)}+D(k)e^{-i(kx-\omega_kt)}\right].
\label{E12}
\end{eqnarray}
We now introduce a vacuum for the Hilbert space, and as usual take the two positive frequency operators $A(k)$ and $B(k)$ to annihilate the right vacuum $|\Omega \rangle$, and the two negative frequency operators $C(k)$ and $D(k)$ to annihilate the left vacuum $ \langle \Omega|$. In addition we require that the vacuum expectation values of the commutators $[C(k),B(k^{\prime})]$ and $[A(k),D(k^{\prime})]$ be given as
\begin{eqnarray}
&&\langle \Omega |[C(k),B(k^{\prime})]|\Omega \rangle=-\langle \Omega |B(k)C(k)|\Omega \rangle\delta(k-k^{\prime})=-f_{BC}(k)\delta(k-k^{\prime}),
\nonumber\\
&&\langle \Omega |[A(k),D(k^{\prime})]|\Omega \rangle=~~\langle \Omega |A(k)D(k)|\Omega \rangle\delta(k-k^{\prime})=~~f_{AD}(k)\delta(k-k^{\prime}).
\label{E13}
\end{eqnarray}
Finally from the consistency of (\ref{E10}) in the form $\langle \Omega |G^{00}(2)|\Omega \rangle/\kappa_2^2=(\hbar/8\pi)\int_{-\infty}^{\infty}dk k[f_{BC}(k)-f_{AD}(k)]=-\rho_{\rm M}$, $\langle \Omega |G^{01}(2)|\Omega \rangle/\kappa_2^2=(\hbar/8\pi)\int_{-\infty}^{\infty}dk \omega_k[f_{BC}(k)-f_{AD}(k)]=0$, we find that the c-number functions $f_{BC}(k)$ and $f_{AD}(k)$ have to obey just the one relation
\begin{equation}
k[f_{BC}(k)-f_{AD}(k)]=4\omega_k=4|k|.
\label{E14}
\end{equation}
With this one quantization condition, we achieve our primary purpose of showing how the quadratically divergent zero-point fluctuations of the gravitational and matter fields mutually cancel each other identically, just as desired. Moreover, through its coupling to the quantized fermionic field the gravitational field commutators are forced to obey (\ref{E14}), with gravity not needing any independent quantization of its own.\footnote{As such, the situation in the gravity case is quite different from that associated with the other interactions. Specifically, in the case of the matter fields one obtains equations of motion by varying the matter action with respect to the matter fields. A canonical quantization then gives the matter energy-momentum tensor $T^{\mu\nu}_{M}$ (the variation of the matter action with respect to the metric) a non-vanishing zero-point contribution. However, for gravity the relevant field is the metric itself. If we define the variation of the gravitational action with respect to the metric to be a quantity $T^{\mu\nu}_{GRAV}$, the gravitational equation of motion is then given by $T^{\mu\nu}_{GRAV}=0$. Then, with $T^{\mu\nu}_{GRAV}$ containing products of fields at the same point, a quantization of the gravitational field would give a zero-point contribution to $T^{\mu\nu}_{GRAV}$ and thus violate the stationarity condition $T^{\mu\nu}_{GRAV}=0$ that $T^{\mu\nu}_{GRAV}$ has to obey. Hence, unlike the matter fields for which there is no constraint on $T^{\mu\nu}_{M}$ in the absence of any coupling of matter to gravity, gravity itself is always coupled to gravity (c.f. the non-linear $G^{\mu\nu}(2)$) and thus cannot be consistently quantized on its own. Moreover, since the infinity in $T^{\mu\nu}_{M}$ is due to the product of two fermion fields at the same point, it would have to be the product of two gravitational fields at the same point (viz. the non-linear $G^{\mu\nu}(2)$) that would have to provide the cancellation.}

\section{Mass Generation and the Cosmological Constant}
\label{S4}

For our purposes here we can take the fermionic action to be of the flat spacetime form $I_{\rm M}=-\int d^2x[i\hbar\bar{\psi}\gamma^{\mu}\partial_{\mu}\psi-(g/2)(\bar{\psi}\psi)^2]$. With the four-Fermi coupling constant $g$ being dimensionless in two dimensions, $I_{\rm M}$ is conformal invariant. Consequently, as well as being covariantly conserved, its energy-momentum tensor $T^{\mu\nu}_{\rm M}=i\hbar\bar{\psi}\gamma^{\mu}\partial^{\nu}\psi -\eta^{\mu\nu}(g/2)[\bar{\psi}\psi]^2$ is traceless in solutions to the equation of motion. In the Nambu-Jona-Lasinio mean-field, Hartree-Fock approximation one looks for self-consistent, translation invariant states $|S\rangle$ in which  
$\langle S|\bar{\psi}\psi | S\rangle=\langle S|\psi^{\dagger}\gamma^0\psi | S\rangle=im/g$ and $\langle S|(\bar{\psi}\psi -im/g)^2| S\rangle=0$. (With our choice of $\eta^{00}=-1$, $\gamma^0$ is pure imaginary.) In such states the fermion equation of motion takes the form $i\hbar\gamma^{\mu}\partial_{\mu}\psi -im\psi=0$ and the mean-field fermion energy-momentum tensor $T^{\mu\nu}_{\rm MF}$  takes the form 
\begin{equation}
\langle S|T^{\mu\nu}_{\rm MF}| S\rangle=\langle S|i\hbar\bar{\psi}\gamma^{\mu}\partial^{\nu}\psi| S\rangle +\frac{m^2}{2g}\eta^{\mu\nu}, 
\label{E15}
\end{equation}
with the mean-field approximation preserving tracelessness. In conformal invariant theories then, we see that, just as noted in \cite{R1}, one can have mass generation without the trace needing to be non-zero. With the emergence of the $(m^2/2g)\eta^{\mu\nu}$ term in (\ref{E15}), we see that dynamical mass generation induces a mean-field cosmological constant term $\Lambda_{\rm MF} =-m^2/2g$, and that with this $\Lambda_{\rm MF}$ we can write $\langle S|T^{\mu\nu}_{\rm MF}| S\rangle$ as
\begin{equation}
\langle S|T^{\mu\nu}_{\rm MF}| S\rangle=(\rho_{\rm MF}+p_{\rm MF})U^{\mu}U^{\nu}+p_{\rm MF}\eta^{\mu\nu} -\Lambda_{\rm MF}\eta^{\mu\nu} ,
\label{E16}
\end{equation}
where
\begin{eqnarray}
\rho_{\rm MF}&=&-\frac{\hbar}{2\pi}\left[K^2+\frac{m^2}{2\hbar^2}+\frac{m^2}{2\hbar^2}{\rm ln}\left(\frac{4 \hbar^2K^2}{m^2}\right)\right],
\nonumber\\
p_{\rm MF}&=&-\frac{\hbar}{2\pi}\left[K^2+\frac{m^2}{2\hbar^2}-\frac{m^2}{2\hbar^2}{\rm ln}\left(\frac{4 \hbar^2K^2}{m^2}\right)\right],\qquad 
\Lambda_{\rm MF}=\frac{m^2}{4\pi\hbar}{\rm ln}\left(\frac{4 \hbar^2K^2}{m^2}\right),
\label{E17}
\end{eqnarray}
and where the expression for $\Lambda_{\rm MF}=-m^2/2g$ is recognized as the gap equation $m=2 \hbar Ke^{\pi\hbar/g}$.

In (\ref{E17}) we see that the mass-independent quadratic divergences in $\rho_{\rm MF} $ and $p_{\rm MF}$ have been augmented by mass-dependent logarithmic ones, while the induced  $\Lambda_{\rm MF}$ is logarithmically divergent (i.e. not finite). However, since $\langle S|T^{\mu\nu}_{\rm MF}| S\rangle$ is traceless, the various terms in (\ref{E17}) obey $p_{\rm MF}-\rho_{\rm MF} -2\Lambda_{\rm MF}=0$, with all the various divergences canceling each other in the trace, just as noted in \cite{R1,R5}. Given this cancellation, we can now use the trace condition to eliminate $\Lambda_{\rm MF}$ and rewrite (\ref{E16}) as 
\begin{equation}
\langle S|T^{\mu\nu}_{\rm MF}| S\rangle=(\rho_{\rm MF}+p_{\rm MF})\left[U^{\mu}U^{\nu}+\frac{1}{2}\eta^{\mu\nu}\right],\qquad
\rho_{\rm MF}+p_{\rm MF}
=-\frac{\hbar}{\pi}\left(K^2+\frac{m^2}{2\hbar^2}\right),
\label{E18}
\end{equation}
with the logarithmic divergences associated with the mass-induced readjustment of $\rho_{MF}$ and $p_{MF}$ having disappeared completely.\footnote{As well as serve to generate fermion masses, the four-Fermi interaction term can also serve to cancel vacuum energy infinities, with a four-Fermi interaction term serving as a vacuum-energy counter-term. Thus, as noted in  P.~D.~Mannheim,  Phys.~Rev.~D {\bf 10}, 3311 (1974); Phys.~Rev.~D {\bf 12}, 1772 (1975);  Nucl.~Phys.~B {\bf 143}, 285 (1978), one could even use such counter-terms in theories such as flat specetime QED. While one does not ordinarily use such four-Fermi counter-terms in flat spacetime QED, once QED is coupled to gravity  one can no longer normal order the vacuum energy away, and the use of four-Fermi counter-terms is then to be preferred.} Finally, to cancel the remaining quadratic divergence and finite part in (\ref{E18}), we proceed just as in the massless fermion case, only with (\ref{E14}) having to be replaced by 
\begin{equation}
k[f_{BC}(k)-f_{AD}(k)]=4\left[(k^2+m^2/\hbar^2)^{1/2}-\frac{m^2}{2\hbar^2(k^2+m^2/\hbar^2)^{1/2}}\right].
\label{E19}
\end{equation}
In (\ref{E19}) we note that even though the gravitational field is massless and still obeys (\ref{E11}), its quantization condition depends on the mass of the fermion, a reflection of  the fact that it is only through the  quantization of the fermionic field that the gravitational field is quantized in the first place.

As such, the above analysis shows how the vacuum contribution to the cosmological constant is completely taken care of by the zero-point contributions. However, there is one further concern that still needs to be addressed, as there is a further contribution to the cosmological constant term matrix element, namely that associated with occupying not just the vacuum $|S\rangle$ but the one-particle excitations as well. In such states $|C\rangle$ the quantity $im(x)/g=\langle C|\bar{\psi}(x)\psi(x)|C\rangle$ can typically acquire a spacetime dependence. However, as noted in \cite{R4}, the spatial dependence will asymptote to the constant vacuum value while the time dependence will redshift. The cosmological term needed for cosmology is thus given not by the vacuum contribution itself but by the spatial departure from it, i.e. by $m^2(x)/2g-m^2/2g$, as redshifted to the current era. Such an effective cosmological term would not at all be constrained to be as large as the vacuum value, but its actual value still needs to be determined.

This paper is based in part on a presentation made by the author at the International Conference on Two Cosmological Models, Universidad Iberoamericana, Mexico City, November 2010. The author wishes to thank Dr. J. Auping and Dr. A. V. Sandoval for the kind hospitality of the conference. The author also wishes to thank Dr. D. J. Gross for helpful comments.

\end{document}